\documentclass[thmsa,onecolumn,a4paper]{article}
\usepackage{amssymb}
\usepackage{sw20jart}
\usepackage{amsmath}
\usepackage{graphicx}

\input{tcilatex}
\begin{document}

\author{Bertrand Chauvineau}
\title{Revisiting the Entangled Relativity gravity framework}
\date{Universit\'{e} C\^{o}te d'Azur, OCA, CNRS, Lagrange, Boulevard de
l'Observatoire, CS 34229, F06304 Nice Cedex 4, France\\
(e-mail : Bertrand.Chauvineau@oca.eu)}
\maketitle

\begin{abstract}
One revisits the Entangled Relativity theory framework, embedding it in a
larger set of theories. From their scalar-tensor reformulation, these
theories admit General Relativity spacetime solutions as solutions with the
same matter content, if this matter obeys an "intrinsic decoupling"
condition. One shows that a set of matter Lagrangians satisfying this
condition is completely characterized by the way they depend on the metric
tensor. One then shows that it is possible to define a perfect fluid
Lagrangian fulfilling the intrinsic decoupling condition up to 1PN order.
Specifying to weak field spherical stars suggests that Entangled Relativity
and its proposed generalization are worthy alternatives to General
Relativity. Indeed, the obtained numerical values of the Eddington
parameters abruptly satisfy the known observational constraints. However,
the non strict compliance of the theory with conservation laws demands to
reconsider the link between observational data and the parameters for a
definitive clarification of this point.
\end{abstract}

\noindent \baselineskip12truept

\noindent \noindent \textbf{I-- Introduction}

\noindent \noindent About a decade ago, O. Minazzoli and collaborators
proposed a new gravity theory, they named Entangled Relativity (ER) [1]-[7].
This theory is purely metric in its basic formulation, in the sense that
gravitation is described by a metric tensor only, alike General Relativity
(GR). However, the Lagrangian from which the theory proceeds was built so
that the strict absence of matter prevents any spacetime theory to emerge,
since the whole action vanishes. In other terms, there is no gravity sector
defined regardless the spacetime matter content (thence the "entangled"
qualifier in the name.) Thence, from the very start, and in deep contrast to
GR, ER doesn't admit solutions with vacuum regions. However, a scalar-tensor
(ST) reformulation of ER is possible, in which such solutions can finally be
contemplated. However, from the point of view of the (initial) purely metric
formulation of the theory, such vacuum states should be interpreted in limit
terms, ie as states with "vanishingly small" matter contents.

The aim of this paper is to revisit the ER framework, and to enlarge it to a
one parameter ($q$) family of theories, we name $q$-ER. The original ER is
recovered in the $q=2$ case. Alike ER, these theories make no sense without
matter (letting apart the $q=0$ case, which is nothing but vacuum GR), and
all of them admit an ST reformulation (which, alike the ST reformulation of
ER, admits solutions with vacuum regions), that differs from the Brans-Dicke
(BD) ST theory [8]. However, it turns out that while all these ST theories
differ each other (and from the BD theory), they nevertheless all identify
with ($\omega =0$)-BD ($0$-BD in brief) in any vacuum region of the
spacetime. Therefore, the differences between the ST reformulation of $q$-ER
(and $0$-BD) \textit{whole spacetime}\ solutions originate in the fact that
these theories differ in the matter filled regions of the spacetime. Thanks
to this, the $q$-ER theories cannot be discarded as a mere consequence that $%
0$-BD\ doesn't success w.r.t. Solar System experiments \noindent \lbrack 9].
Besides, it is possible to show that GR solutions are also $q$-ER
(particular) solutions when a specific condition, named intrinsic decoupling
(ID) after [5], is filled by matter. Involving the Euler theorem, a set of
matter Lagrangians can be identified that automatically satisfy the ID
requirement. However, this sufficient condition for ID is not necessary:
matter fields exist whose Lagrangian don't belong the previous set, but
nonetheless comply with ID thanks to the implementation of the full set of
field equations (ie, involving the on-shell expression of the Lagrangian).

Having astronomical applications in mind, one then focuses our attention to
perfect fluid (PF) made sources. Our starting point is the Brown PF
Lagrangian [10] and a generalization with a divergence term added, that
involves a constant parameter $h$. All these Lagrangians are equivalent in
the GR context, but aren't in the $q$-ER one, the Lagrangian itself being
involved in the field equations. In the $q$-ER framework, these Lagrangian
don't strictly satisfy the ID requirement, apart from the case of a very
specific $\left( q,h\right) $ duet. Nonetheless, it is shown that whatever $%
q $, a specific ($q$ dependent) choice of $h$ yields a Lagrangian that
satisfies ID up to 1PN order (PN for Post-Newtonian). Implementing the
resulting Brown like Lagrangian, one obtains a numerical value of the $%
\gamma _{Edd}$ Eddington parameter for a spherical star. The departure from $%
1$ of this value is about one order of magnitude below the experimental
constraints to date, which suggests that $q$-ER theories could be worth
alternatives to GR. However, it is also stressed that such a claim would
first require a careful analysis of the way test PF balls behave in $q$-ER
gravity.

\qquad

\noindent \noindent \textit{Outline of the paper}

\noindent \noindent The paper is organized as follows. In Sect. II is first
reminded the ER action and equations, through their proposed $q$-ER
generalization. The ST reformulation is detailed, and the link and
differences between the two formulations are emphasized. It is then shown
that a specific set of matter Lagrangians naturally satisfies the ID
requirement. A one parameter family of possible PF Lagrangians for GR is
proposed in Sect. III, and the form that allows the existence of close to GR 
$q$-ER\ solutions (ie: the ID requirement is not exactly, but only "nearly"
satisfied) is fixed. Sect. IV is a digression towards the induced Poisson
equation. Sect. V is devoted to spherisymmetric solutions describing the
external field of a static star. The $\gamma _{Edd}$ Eddington parameter is
obtained, and the ability of the previous PF Lagrangian fixing to make $q$%
-ER an experimentally viable alternative to GR is discussed. Sect. VI
concludes the paper.

\qquad

For calculations to come, one chooses the $\left( -,+,+,+\right) $ signature
for the metric.

\qquad

\noindent \noindent \textbf{II -- Field equations of a general power law
Lagrangian}

\noindent \noindent Let us consider the ER inspired action\footnote{%
The absolute value only for the possibility to consider non integer values
of $q$. If only integer values are considered, the action still makes sense
with the absolute value removed, at least when only classical gravity issues
are into consideration.}%
\begin{equation}
S_{q}\left[ g_{\ast \ast };\Psi \right] =\int d^{4}x\sqrt{-g}R\left\vert 
\frac{L_{m}}{R}\right\vert ^{q}=\int d^{4}x\sqrt{-g}R\left( s\frac{L_{m}}{R}%
\right) ^{q}  \label{q-ER action}
\end{equation}%
where $R$ is the metric scalar curvature, $L_{m}$ a matter fields
(collectively represented by $\Psi $) dependent scalar function (named
matter Lagrangian), $q$ a real number and $s$ the sign of $\frac{L_{m}}{R}$.
Apart from the matter fields, the action just depends on the metric tensor
field. In this sense, the theory resulting from (\ref{q-ER action}) is a
purely tensor gravity theory. For having a scalar curvature dependent
action, it is natural to demand $q\neq 1$. One defines the matter stress
tensor from $L_{m}$ following the usual procedure%
\begin{equation}
\delta \left( \sqrt{-g}L_{m}\right) =-\frac{1}{2}\sqrt{-g}T_{ab}\delta
g^{ab}=\frac{1}{2}\sqrt{-g}T^{ab}\delta g_{ab}.  \label{stress tensor def}
\end{equation}%
Varying (\ref{q-ER action}) w.r.t. the metric, one obtains the "Einstein $q$%
-ER equation"%
\begin{equation}
E_{ab}=\frac{q}{2\left( 1-q\right) }\frac{R}{L_{m}}T_{ab}+\frac{1}{H}\left(
\nabla _{b}\partial _{a}-g_{ab}\square \right) H\text{ \ \ with \ \ }%
H=\left\vert \frac{L_{m}}{R}\right\vert ^{q}.  \label{q-ER field eq}
\end{equation}%
where $E_{ab}\equiv R_{ab}-\frac{1}{2}Rg_{ab}$\ is the Einstein tensor. One
will hereafter name $q$-ER the corresponding theory (refering to the
original ER theory, that corresponds to the $q=2$\ case). Let us stress that
the "scalar field" $H$ is only a by-product of the dynamical fields entering
the action (\ref{q-ER action}). For the factor in front of the stress tensor
in (\ref{q-ER field eq}) to be positive (like in GR), let us only consider
solutions of $q$-ER satisfying%
\begin{equation}
\frac{q}{2\left( 1-q\right) }\frac{R}{L_{m}}\geq 0\text{ \ \ \TEXTsymbol{<}%
---\TEXTsymbol{>} \ \ }sq\left( 1-q\right) \geq 0.  \label{positivity cond}
\end{equation}%
(We will be back to this point later on, when a Lagrangian for the PF would
have been fixed.) The trace of (\ref{q-ER field eq}) yields%
\begin{equation}
\frac{3}{H}\square H=\frac{R}{L_{m}}\left( L_{m}-\frac{q}{2\left( q-1\right) 
}T\right)  \label{q-ER scalar eq}
\end{equation}%
that one will name the "scalar $q$-ER equation"\footnote{%
Note the difference with the BD theory: the BD scalar equation is
independent on the Einstein BD equation, while the $q$-ER scalar equation is
included in the Einstein $q$-ER one. (This is due to the fact that $H$ is
not a fundamental gravitational field entering the action, while the scalar
field of the BD theory is. This is very similar to what happens when the $%
f\left( R\right) $ theory is reinterpreted as a specific BD one.)}. Now, the
divergence of (\ref{q-ER field eq}) yields, after some calculations, and
re-using (\ref{q-ER field eq})%
\begin{equation}
\nabla _{b}\left( H^{1-\frac{1}{q}}T_{a}^{b}\right) =L_{m}\partial
_{a}\left( H^{1-\frac{1}{q}}\right) \text{ \ \ \TEXTsymbol{<}---\TEXTsymbol{>%
} \ \ }\nabla _{b}T_{a}^{b}=\frac{q-1}{q}\left( L_{m}\delta
_{a}^{b}-T_{a}^{b}\right) \partial _{b}\ln H
\label{q-ER stress tensor non conserv}
\end{equation}%
which shows that the stress tensor is generically \textit{not} conserved in $%
q$-ER gravity.

\qquad

\noindent \noindent \textbf{II.1 -- Scalar renormalization and effective ST
action}

\noindent \noindent It is worth renormalizing the scalar by defining $\Phi $
($>0$) as%
\begin{equation}
\frac{1}{8\pi }\Phi ^{\frac{1}{q}}=\frac{2\left( 1-q\right) }{q}\frac{L_{m}}{%
R}  \label{scalar renorm}
\end{equation}%
which is possible thanks to (\ref{positivity cond}). Let us remark that this
scalar (or $H$) vanishes when $L_{m}/R$ does in the $q>0$ case, while it
diverges in the $q<0$\ one. The (\ref{q-ER field eq}), (\ref{q-ER scalar eq}%
) and (\ref{q-ER stress tensor non conserv}) equations rewrite 
\begin{subequations}
\begin{eqnarray}
E_{ab} &=&8\pi \Phi ^{-\frac{1}{q}}T_{ab}+\frac{1}{\Phi }\left( \nabla
_{b}\partial _{a}-g_{ab}\square \right) \Phi  \label{q-ER renorm field eq} \\
\square \Phi &=&\frac{8\pi }{3}\Phi ^{1-\frac{1}{q}}\left( T-2\frac{q-1}{q}%
L_{m}\right)  \label{q-ER renorm scalar eq} \\
\nabla _{b}\left( \Phi ^{1-\frac{1}{q}}T_{a}^{b}\right) &=&L_{m}\partial
_{a}\left( \Phi ^{1-\frac{1}{q}}\right)
\label{q-ER renorm stress tensor non conserv}
\end{eqnarray}%
with 
\end{subequations}
\begin{equation}
\Phi =\left( 16\pi \frac{1-q}{q}\frac{L_{m}}{R}\right) ^{q}=\left( 16\pi s%
\frac{1-q}{q}\right) ^{q}H  \label{q-ER Phi def}
\end{equation}%
that are the (redundant) equations of the $q$-ER theory, but in its ST
reformulation. For reasons we will discuss later, let us name $q$-EST (EST
for Entangled ST) the theory described by the equations (\ref{q-ER renorm
field eq})-(\ref{q-ER renorm stress tensor non conserv}), considered
discarding the initial $q$-ER standpoint they are coming from. As said
before, the $1$-ER theory should be discarded, being not a true gravity
theory. Let us remark that, concomitantly, the $q=1$ value of the
renormalized scalar (\ref{q-ER Phi def}) vanishes. Nonetheless, it could be
worth to remark that the $q=1$ version of the $q$-EST equations (considered
independently) are exactly the $0$-BD theory equations\footnote{%
One could also remark that if the $q=1$ case is considered not abruptly but
as a $q\longrightarrow 1$ limit process (that makes sense since the theory
is well defined at each step), a spacetime theory emerges (albeit a kind of
renormalized interpretation of the scalar) that is \textit{not} the
Minkowski spacetime, but a theory that involves gravity like effects,
despite the fact that (\ref{q-ER action}) reduces to a purely matter action
(with no curvature contribution involved, ie with not gravity theory
proceeding from the $q=1$\ version of the (\ref{q-ER action}) action
itself). However, the emergent gravity theory, ie $0$-BD, is experimentally
discarded.}.

On the other hand, the (\ref{q-ER renorm field eq})-(\ref{q-ER renorm stress
tensor non conserv}) $q$-EST equations for $q\neq 1$ differ from $0$-BD by
the following points:

-- $\Phi $ appears in front of $T_{ab}$ in (\ref{q-ER renorm field eq})\
with a power that differs from $-1$

-- the scalar is not $T$ only sourced, the sourcing also involving the
matter Lagrangian itself (besides some power of the scalar itself as a
global source factor)

-- the stress tensor is not conserved.

All these points also distinguish the $q$-EST theories each other, these
differences with $0$-BD happening in a$\ q$ dependent way.

\qquad

\noindent \noindent Incidentally, one can show that the ($q$-EST) action%
\begin{equation}
S\left[ g_{\ast \ast },\Phi ;\Psi \right] =\int d^{4}x\sqrt{-g}\left( \Phi
R+16\pi \Phi ^{1-\frac{1}{q}}L_{m}\right)  \label{q-ER renorm eff action}
\end{equation}%
returns the Einstein and scalar equations (\ref{q-ER renorm field eq}) and (%
\ref{q-ER renorm scalar eq}), from which the $\Phi $ expression\ in terms of 
$L_{m}/R$\ in (\ref{q-ER Phi def}) can be recovered. It returns the original
ER ST action [1] for $q=2$.

\qquad

\noindent \noindent \textbf{II.2 -- Is it finally that illegitimate
considering vacuum }$q$\textbf{-ER, or }$q$\textbf{-EST, solutions?}

\noindent \noindent The $q$-ER purely metric theory has been built in such a
way that considering (partly or fully) vacuum solutions has no meaning.
However, both the $q$-EST action (\ref{q-ER renorm eff action}) and field
equation (\ref{q-ER renorm field eq}) \textit{do make sense} in vacuum, and $%
q$-EST indeed admits vacuum solutions. Is there a mismatch here? The point
is that the $q$-ER versus $q$-EST "equivalence" should be more precisely
requalified as a \textit{quasi}-equivalence, in the sense that it just
concerns solutions having no vacuum regions. This means that while any $q$%
-ER solution is a $q$-EST one, a $q$-EST solution is a $q$-ER one only if it
doesn't exhibit vacuum regions, and only in this case. If it does, the $q$%
-EST solution should to be just regarded as the limit of a $q$-ER solution
series, from the $q$-ER standpoint.

Now, if the $q$-ER theory turns out to have something to do with our true
Universe, it has to be able to describe setups involving close to vacuum
spacetime regions, since our Universe exhibits many such cases. This can be
made in both $q$-ER and $q$-EST frameworks. But if $q$-EST close to vacuum
regions asymptotically behave like $q$-EST strictly vacuum regions, as it
has been checked on several specific $2$-ER solutions (see for instance
[11]), it should be better to directly solve vacuum $q$-EST equations, which
are simpler thanks to vanishing terms (this is also the point of view that
has been adopted in [19]). In this frame of mind, we will (1) consider such $%
q$-EST solutions in the following, and (2) admit that these $q$-EST
solutions appropriately represent the $q$-ER solutions that describe the
same physical setup, but formulated in $q$-ER terms (ie: with some regions
filled with vanishingly small matter content).

In the following, one will generically refer to $q$-ER when general
properties of the theory are inferred, independently of some specific
solution, while to $q$-EST when solutions explicitely involving vacuum
spacetime regions are into consideration.

\qquad

\noindent \noindent \textbf{II.3 -- Vacuum }$q$\textbf{--EST equations}

\noindent \noindent In vacuum regions of the spacetime, (\ref{q-ER renorm
stress tensor non conserv}) is trivial, while (\ref{q-ER renorm field eq})
and (\ref{q-ER renorm scalar eq}) simplify into 
\begin{subequations}
\begin{eqnarray}
R_{ab} &=&\frac{1}{\Phi }\nabla _{b}\partial _{a}\Phi  \label{vac ER EE} \\
\square \Phi &=&0.  \label{vac ER scalar}
\end{eqnarray}%
These equations are nothing but the vacuum $0$-BD equations. In this sense,
all the $q$-EST theories merge into a unique theory, which is the $0$-BD
one, in vacuum. However, claiming that $q$-EST (thence $q$-ER) theories
should be rejected from the fact that $0$-BD doesn't pass the experimental
tests would be premature. Indeed, the link between the parameters describing
the external field of a source made of matter explicitely involves the field
equations \textit{inside} the source\footnote{%
With some regularity conditions (at the center in the spherical case, for
instance) to be satisfied.}. The latter differing for $0$-BD and $q$-EST (ie 
$q$-ER) theories, the link between the parameters of the external ($0$-BD)
solution is \textit{not} the same in all these theories. This results in
different values of the $\gamma _{Edd}$ Eddington parameter, for instance%
\footnote{%
Having in mind that the $\beta _{Edd}$ Eddington parameter is always equal
to $1$ in BD gravity.}. This will be illustrated in an explicit example in
Sect. V.

On the other hand, (\ref{vac ER EE}-\ref{vac ER scalar}) show that $q$-EST
admits all the \textit{fully} vacuum $0$-BD solutions (with singularities or
not). Specifying to the $\partial \Phi =0$\ case, this encompasses any
vacuum GR solution (Schwarzschild, Kerr, Kasner, ... and bags of known exact
solutions [12]). This also includes all the (stationary and dynamic) multi
black hole GR (analytic and numerical) solutions. Beyond vacuum GR, this
also encompasses the wormhole like and naked singularity Brans Class I $0$%
-BD solutions \noindent \lbrack 13].

\qquad

\noindent \noindent \textbf{II.4 -- About test particles motion in }$q$%
\textbf{-ER and }$q$\textbf{-EST}

\noindent \noindent In a spacetime exhibiting vacuum regions, the $q$-EST
metric outside matter is exactly $0$-BD. One could be tempted to conclude
that photons and (weakly massive enough) massive bodies should move
following $0$-BD geodesics in such regions, as they do in $0$-BD gravity.
This issue deserves a more careful discussion, based on the way test
particle motions are mastered by the spacetime geometry.

In usual gravity, like GR, BD and ST gravity, matter Lagrangians enter the
action additively. Let us consider that in some spacetime region, only test
matter is present. This can be formalized writting the total Lagrangian in
the form 
\end{subequations}
\begin{equation}
L\left[ g_{\ast \ast },\Xi ;\psi \right] =L_{grav}\left[ g_{\ast \ast },\Xi %
\right] +L_{m}\left[ g_{\ast \ast };\psi \right] =L_{grav}\left[ g_{\ast
\ast },\Xi \right] +\varepsilon l_{m}\left[ g_{\ast \ast };\psi \right]
\label{usual theories}
\end{equation}%
where the $L_{grav}$ sector depends on the metric and other gravitational
fields (if so), collectively represented by $\Xi $.\ On the other hand, only
the metric enters the matter sector (in Jordan representation), besides the $%
\psi $\ matter fields. Rewriting $L_{m}$\ in the $\varepsilon l_{m}$ form,
involving a vanishingly small $\varepsilon $\ quantity,\ formalizes the
"test status" of the $\psi $ fields, $l_{m}$\ being $\varepsilon $
independent. The variation w.r.t. the metrics returns the vacuum field
equations, the $\psi $ stress tensor being vanishingly damped by the $%
\varepsilon $ factor. On the other hand, $L_{grav}$ is not involved in the
matter field equations, in which $\varepsilon $ simplifies as a global
factor. This yields unambiguous $\varepsilon $ independent matter equations.

What happens when switching from (\ref{usual theories}) like theories to (%
\ref{q-ER action}) like ones? Nearly vacuum spacetime regions are endowed
with a $L_{m}$ matter Lagrangian, that one can represent in the $\varepsilon
l_{m}\left[ g_{\ast \ast };\psi \right] $ form, alike we did in the (\ref%
{usual theories}) case. From (\ref{stress tensor def}), the stress tensor is
endowed with the same $\varepsilon $\ factor as $L_{m}$. Accordingly, let us
write it $T_{ab}=\varepsilon t_{ab}$, where $t_{ab}$\ is $\varepsilon $
independent. The (\ref{q-ER field eq}) Einstein equation is unchanged, with
just $L_{m}$\ and $T_{ab}$\ changed for $l_{m}$\ and $t_{ab}$\ respectively,
after an obvious $\varepsilon $\ simplification. It is then unaffected by
the $\varepsilon \longrightarrow 0$\ process. Since (\ref{q-ER scalar eq})
and (\ref{q-ER renorm stress tensor non conserv}) proceed from (\ref{q-ER
field eq}), they are no more affected, modulo the previous substitutions, as
it can be checked directly. One can be puzzled by the fact that the scalar $H
$ vanishes or diverges in the $\varepsilon \longrightarrow 0$\ process,
depending on the $q$'s sign. However this by no means impacts the effective
equations. Using the $H$ expression in (\ref{q-ER field eq}), (\ref{q-ER
renorm stress tensor non conserv}) yields%
\begin{equation}
\nabla _{b}t_{a}^{b}=\left( 1-\frac{1}{q}\right) \left( l_{m}\delta
_{a}^{b}-t_{a}^{b}\right) \partial _{b}\ln H=\left( q-1\right) \left(
l_{m}\delta _{a}^{b}-t_{a}^{b}\right) \partial _{b}\ln \left\vert \frac{l_{m}%
}{R}\right\vert   \label{test matter non conserv}
\end{equation}%
that explicits the $\varepsilon $ independence of test matter motion.
Thence, (\ref{test matter non conserv}) shows that test matter doesn't obey
conservation laws, despite the fact that the $q$-EST field equations reduce
to the $0$-BD ones.

\qquad

\noindent \noindent \textbf{II.5 -- Matter field equations and ID}

\noindent \noindent Let us now focuse on the matter field equations. The
Lagrange equation for the matter fields reads, from (\ref{q-ER action}) 
\begin{equation}
\partial _{a}\left[ \frac{\partial \left( \sqrt{-g}R\left\vert \frac{L_{m}}{R%
}\right\vert ^{q}\right) }{\partial \left( \partial _{a}\Psi \right) }\right]
=\frac{\partial \left( \sqrt{-g}R\left\vert \frac{L_{m}}{R}\right\vert
^{q}\right) }{\partial \Psi }.  \label{q-ER matter eq}
\end{equation}%
This identifies with the usual matter field equations if $R\left\vert \frac{%
L_{m}}{R}\right\vert ^{q}\propto L_{m}$, ie if\ $R\propto L_{m}$, which
means that the $H$, or $\Phi $, scalar is constant. Accordingly, the stress
tensor conservation is recovered from (\ref{q-ER renorm stress tensor non
conserv}). The Einstein equation (\ref{q-ER renorm field eq}) reduces to the
usual GR equation, with the Newton constant $G=\Phi ^{-\frac{1}{q}}$. The
scalar equation (\ref{q-ER renorm scalar eq}) yields%
\begin{equation}
L_{m}=\frac{q}{2\left( q-1\right) }T  \label{ID cond 1}
\end{equation}%
a condition that we will name ID, in accordance with [5] terminology.

Let us point out that (\ref{ID cond 1}) is to be written concomitantly with
the other field equations, ie the on-shell expression of $L_{m}$\ is
involved. However, it is worth pointing out that demanding $L_{m}$ to
satisfy (\ref{ID cond 1}) \textit{regardless} of the other matter equations
(off-shell $L_{m}$) allows to characterize a set of Lagrangians complying
with ID.\ Indeed, from the stress tensor definition (\ref{stress tensor def}%
), (\ref{ID cond 1}) rewrites, specifying to $L_{m}$'s that depend on the
metric but not on its derivatives (a usual assumption on matter-metric
couplings)%
\begin{equation}
g^{ab}\frac{\partial D_{m}}{\partial g^{ab}}=\left( \frac{1}{q}-1\right)
D_{m}\text{ \ \ where \ \ }D_{m}=\sqrt{-g}L_{m}.  \label{ID cond 2}
\end{equation}%
It turns out that the Euler theorem completely and unambiguously fixes the
set of matter Lagrangians satisfying (\ref{ID cond 2}): these are
Lagrangians such that $D_{m}$\ is $\left( \frac{1}{q}-1\right) $-$g^{\ast
\ast }$-homogeneous (ie: homogeneous functions of degree $\left( \frac{1}{q}%
-1\right) $ of the contravariant metric components). Thence, considering
only 4-dimensional spacetimes, the set includes all the $L_{m}$'s that are $%
\left( \frac{1}{q}+1\right) $-$g^{\ast \ast }$-homogeneous and only these
Lagrangians. All off-shell Lagrangians satisfying (\ref{ID cond 1}) also do
when the other matter equations are implemented (on-shell expression of $%
L_{m}$). On the other hand, it is possible that specific matter fields
described by a Lagrangians not belonging to this set nevertheless satisfy
the on-shell version of (\ref{ID cond 1}), thanks to the implementation of
the other field equations.

Reciprocally, the ID condition (\ref{ID cond 1}) yields, from (\ref{q-ER
scalar eq}) or (\ref{q-ER renorm scalar eq})%
\begin{equation}
\square H=0\text{ \ \ or \ \ }\square \Phi =0  \label{ID cond 3}
\end{equation}%
in both vacuum and matter. This ensures $T_{\ast \ast }$-filled GR solutions
to be also $q$-EST ones, since GR solutions have constant $\Phi $. Let us
however point out that, even under ID, the theory a priori also admits non
GR solutions, (\ref{ID cond 3}) also admiting non constant solutions in the
general case.

\qquad

\noindent \noindent \textbf{II.6 -- Conformal transformations}

\noindent \noindent Let us consider two conformally related metrics%
\begin{equation}
\widetilde{g}_{ab}=\Omega ^{2}g_{ab}  \label{conf metrics}
\end{equation}%
where $g_{\ast \ast }$ is a $q$-ER solution with the $L_{m}$ matter
Lagrangian, and $\Omega $\ a scalar field. The spacetime being
4-dimensional, the scalar curvatures are related by [14]%
\begin{equation}
\Omega ^{2}\widetilde{R}=R-6\frac{\square \Omega }{\Omega }\text{ \ \ 
\TEXTsymbol{<}---\TEXTsymbol{>} \ \ }\Omega ^{-2}R=\widetilde{R}-6\Omega 
\widetilde{\square }\left( \frac{1}{\Omega }\right)  \label{conf scalar curv}
\end{equation}%
where the tilde on the $\widetilde{\partial }^{\ast }$\ and $\widetilde{%
\square }$ operators means that they are defined from $\partial _{\ast }$
with the help of the tilded metric. Using (\ref{conf scalar curv}), the
action integral in (\ref{q-ER action}) can be rewritten%
\begin{equation}
\int d^{4}x\sqrt{-g}R\left\vert \frac{L_{m}\left( g^{\ast \ast },\Psi
\right) }{R}\right\vert ^{q}=\int d^{4}x\sqrt{-\widetilde{g}}\left[ 
\widetilde{R}-6\Omega \widetilde{\square }\left( \frac{1}{\Omega }\right) %
\right] \left\vert \frac{\Omega ^{-2\left( \frac{1}{q}+1\right) }L_{m}\left(
g^{\ast \ast },\Psi \right) }{\widetilde{R}-6\Omega \widetilde{\square }%
\left( \frac{1}{\Omega }\right) }\right\vert ^{q}
\label{ER int action with conf metric}
\end{equation}%
Now, if $L_{m}$ is $\left( \frac{1}{q}+1\right) $-$g^{\ast \ast }$%
-homogeneous, $H$ satisfies (\ref{ID cond 3}). Remarking that $\square
\Omega =0$\ is equivalent to $\widetilde{\square }\left( \frac{1}{\Omega }%
\right) =0$ (thanks to the spacetime 4-dimensionality), one has%
\begin{equation}
\Omega =H\text{ \ \ ---\TEXTsymbol{>} \ \ }\int d^{4}x\sqrt{-g}R\left\vert 
\frac{L_{m}\left( g^{\ast \ast },\Psi \right) }{R}\right\vert ^{q}=\int
d^{4}x\sqrt{-\widetilde{g}}\widetilde{R}\left\vert \frac{L_{m}\left( 
\widetilde{g}^{\ast \ast },\Psi \right) }{\widetilde{R}}\right\vert ^{q}
\label{ID versus conf transfo}
\end{equation}%
where the $L_{m}$'s $\left( \frac{1}{q}+1\right) $-$g^{\ast \ast }$%
-homogeneity\ has been used. Therefore, if $g_{\ast \ast }$ is a $\Psi $
filled $q$-ER solution, $\widetilde{g}_{\ast \ast }=H^{2}g_{ab}$ is another
one. The "tilde scalar" reads%
\begin{equation}
\widetilde{H}=\left\vert \frac{L_{m}\left( \widetilde{g}^{\ast \ast },\Psi
\right) }{\widetilde{R}}\right\vert ^{q}=\left\vert \frac{L_{m}\left(
H^{-2}g^{\ast \ast },\Psi \right) }{H^{-2}R}\right\vert
^{q}=H^{-2}\left\vert \frac{L_{m}\left( g^{\ast \ast },\Psi \right) }{R}%
\right\vert ^{q}=\frac{1}{H}.  \label{conf scalar}
\end{equation}%
In short terms, if $\left( g_{\ast \ast },H,\Psi \right) $ is a solution, $%
\left( H^{2}g_{\ast \ast },H^{-1},\Psi \right) $ is another one. This
duality shows that in the case of $\left( \frac{1}{q}+1\right) $-$g^{\ast
\ast }$-homogeneous Lagrangians, the solutions go by pairs. However, these
two solutions are the same if $\left( g_{\ast \ast },\Psi \right) $ is a GR
one, since $H$ is constant then.

\qquad

\noindent \noindent \textbf{III -- Matter Lagrangians}

\noindent \noindent In this section, one first reviews the most currently
used Lagrangians. Thence, motivated by its many astronomical applications,
one gives special attention to the more complex PF case.

\qquad

\noindent \noindent \textbf{III.1 -- Free particles and dust}

\noindent \noindent The Lagrangian of a free particle of mass $m$ reads,
parametrizing its orbit $X^{i}\left( t\right) $ with a coordinate time%
\begin{equation}
L_{part}\left( g^{\ast \ast };X^{i}\left( t\right) \right) =-\frac{m}{\sqrt{%
-g}}\sqrt{-g_{ab}\frac{dx^{a}}{dt}\frac{dx^{b}}{dt}}\delta ^{3}\left(
x^{i}-X^{i}\left( t\right) \right) .  \label{free part Lagr}
\end{equation}%
It is clearly a $\left( -\frac{3}{2}\right) $-$g_{\ast \ast }$-homogeneous
functional, ie a $\frac{3}{2}$-$g^{\ast \ast }$-homogeneous one. Therefore,
it satisfies the ID requirement for $q=2$, ie for the original ER theory,
whatever the particle's motion. This is also the case for dust fields, dust
being also a specific case of PFs, made of non interacting free particles.
The associated free particle stress tensor reads%
\begin{equation}
T^{ab}=\frac{m}{\sqrt{-g}u^{0}}\frac{dx^{a}}{d\tau }\frac{dx^{b}}{d\tau }%
\delta ^{3}\left( x^{i}-X^{i}\left( t\right) \right) \text{ \ \ ---%
\TEXTsymbol{>} \ \ }T=-\frac{m}{\sqrt{-g}u^{0}}\delta ^{3}\left(
x^{i}-X^{i}\left( t\right) \right) .  \label{free part stress tensor}
\end{equation}%
Therefore%
\begin{equation}
L_{part}-\frac{q}{2\left( q-1\right) }T=-\frac{q-2}{2\left( q-1\right) }%
\frac{m}{\sqrt{-g}u^{0}}\delta ^{3}\left( x^{i}-X^{i}\left( t\right) \right)
.  \label{free part ID test}
\end{equation}%
Thence, one sees that for $q\neq 2$, the Lagrangian (\ref{free part Lagr})
cannot satisfy the ID requirement. The ID requirement is then satisfied for $%
q=2$, and for this $q$ value only.

\qquad

\noindent \noindent \textbf{III.2 -- Electrovacuum}

\noindent \noindent The electrovacuum Lagrangian%
\begin{equation}
L_{em}\left( g^{\ast \ast };A_{\ast }\right) =g^{ac}g^{be}F_{ab}F_{ce}\text{
\ \ with \ \ }F_{ab}=\partial _{a}A_{b}-\partial _{b}A_{a}  \label{em Lagr}
\end{equation}%
is $2$-$g^{\ast \ast }$-homogeneous. Therefore, it is $\left( \frac{1}{q}%
+1\right) $-$g^{\ast \ast }$-homogeneous if and only if $q=1$, ie for the
discarded case, as argued previously. The associated stress tensor reads%
\begin{equation}
T_{ab}=-4g^{ce}F_{ac}F_{be}+g_{ab}g^{pq}g^{ce}F_{pc}F_{qe}\text{ \ \ ---%
\TEXTsymbol{>} \ \ }T=0.  \label{em stress tensor}
\end{equation}%
Therefore%
\begin{equation}
L_{em}-\frac{q}{2\left( q-1\right) }T=L_{em}=2g^{ac}g^{be}F_{ab}\partial
_{c}A_{e}.  \label{em ID test}
\end{equation}%
It turns out that, whatever $q$, the on-shell expression of (\ref{em ID test}%
) vanishes in the geometric optics approximation (GOA), as shown in the
Appendix.

\qquad

\noindent \noindent \textbf{III.3 -- Scalar fields}

\noindent \noindent The scalar Lagrangian reads%
\begin{equation}
L_{sc}\left( g^{\ast \ast };\Phi \right) =g^{ab}\partial _{a}\Phi \partial
_{b}\Phi +m^{2}\Phi ^{2}.  \label{sc Lagr}
\end{equation}%
In the massless case, (\ref{sc Lagr}) is $1$-$g^{\ast \ast }$-homogeneous.
Therefore, it cannot be $\left( \frac{1}{q}+1\right) $-$g^{\ast \ast }$%
-homogeneous whatever the finite value of $q$. In the massive case, (\ref{sc
Lagr}), the mass term even breaks the Lagrangian $g^{\ast \ast }$%
-homogeneity. The associated stress tensor reads%
\begin{equation}
T_{ab}=-2\partial _{a}\Phi \partial _{b}\Phi +g_{ab}\left[ \left( \partial
\Phi \right) ^{2}+m^{2}\Phi ^{2}\right] \text{ \ \ ---\TEXTsymbol{>} \ \ }%
T=2\left( \partial \Phi \right) ^{2}+4m^{2}\Phi ^{2}.
\label{sc stress tensor}
\end{equation}%
Therefore%
\begin{equation}
L_{sc}-\frac{q}{2\left( q-1\right) }T=-\frac{1}{q-1}\left( \partial \Phi
\right) ^{2}-\frac{q+1}{q-1}m^{2}\Phi ^{2}.  \label{sc ID test}
\end{equation}%
In the massless case, the on-shell version of (\ref{sc ID test}) vanishes
only for scalar fields having radiative like gradients. This is also the
case for massive fields in the $q=-1$ theory. In the other cases, ID
requires the (on-shell expression of the) scalar gradient to satisfy%
\begin{equation}
\left( \partial \ln \Phi \right) ^{2}=-\left( q+1\right) m^{2}.
\label{sc ID test 2}
\end{equation}

\qquad

\noindent \noindent \textbf{III.4 -- Perfect fluids}

\noindent \noindent Considering astronomical applications of gravity
theories (stars, planetary systems, radiative area cosmology, ...), the most
considered matter content is PF with non vanishing pressure. Let us then now
focuse on the Lagrangian description of general PFs.

\qquad

\noindent \noindent \textbf{III.4.a -- The Brown PF Lagrangian for GR}

\noindent \noindent The PF Lagrangian is not as clearly fixed as the ones
reminded in the previous sub-sections. Nevertheless, discarding any
fundamental justification but just demanding it to return the PF stress
tensor, the Brown's Lagrangian [10] is frequently used. It reads%
\begin{equation}
L_{PF}\left( g^{ab};j^{a},\phi ,s,\theta ,\alpha ^{A},\beta _{A}\right)
=-\epsilon \left( n\equiv \sqrt{-g_{ab}j^{a}j^{b}},s\right) +j^{a}\left(
\partial _{a}\phi +s\partial _{a}\theta +\beta _{A}\partial _{a}\alpha
^{A}\right)  \label{Brown PF Lagr}
\end{equation}%
where $n$ and $s$\ are respectively the so called fluid's "rest mass
density" (or "particle number density", or "baryonic density") and the
"entropy per particle". If the $A$ index runs from $1$ to $3$, the $\alpha
^{A}$'s can be interpreted as Lagrangian coordinates of the fluid's
particles [10]. Note that $\epsilon $, that is to be interpreted as the PF's
energy density, depends on the matter vector field $j$\ only through its
modulous $n$. The $j$'s normalized part $u$, defined by 
\begin{equation}
j^{a}=nu^{a}  \label{quadri-velo_def}
\end{equation}%
is to be interpreted as the PF's four-velocity. With these physical
interpretations, (\ref{Brown PF Lagr}) yields, in the GR (additive
Lagrangians) framework, the PF's Euler and baryonic conservation equations.

Indeed, the full set of GR equations derived from (\ref{Brown PF Lagr}) can
be written 
\begin{subequations}
\begin{gather}
E_{ab}=\frac{1}{2}\left[ \left( \epsilon +n^{2}\partial _{n}\left( \frac{%
\epsilon }{n}\right) \right) u_{a}u_{b}+n^{2}\partial _{n}\left( \frac{%
\epsilon }{n}\right) g_{ab}\right]  \label{1} \\
\left( \partial _{n}\epsilon \right) u_{a}=-\partial _{a}\phi -s\partial
_{a}\theta -\beta \partial _{a}\alpha ^{A}  \label{2} \\
\partial _{s}\epsilon =nu^{a}\partial _{a}\theta  \label{3} \\
\nabla _{a}\left( nu^{a}\right) =0  \label{4} \\
u^{a}\partial _{a}s=0  \label{5} \\
u^{a}\partial _{a}\alpha ^{A}=0  \label{6} \\
u^{a}\partial _{a}\beta _{A}=0.  \label{7}
\end{gather}%
Let us remark that contracting (\ref{2}) with $u^{a}$\ yields, using (\ref{5}%
) and (\ref{6}) 
\end{subequations}
\begin{equation}
\partial _{n}\epsilon =u^{a}\partial _{a}\left( \phi +s\theta \right) .
\label{(2) contracted}
\end{equation}%
The (\ref{1}) equation rewrites then 
\begin{equation}
E_{ab}=\frac{1}{2}T_{ab}\text{ \ \ with \ \ }\left\{ 
\begin{array}{l}
T_{ab}\equiv \left( \epsilon +P\right) u_{a}u_{b}+Pg_{ab} \\ 
P\equiv n^{2}\partial _{n}\left( \frac{\epsilon }{n}\right)%
\end{array}%
\right.  \label{GR Einstein eq}
\end{equation}%
where $T_{ab}$ is the PF stress tensor, which justifies the previous $n$, $%
\epsilon $ and $u$ interpretations (as well as $P$\ as the PF's pressure).
From (\ref{2}), the fluid's four-velocity reads%
\begin{equation}
u_{a}=-\frac{\partial _{a}\phi +s\partial _{a}\theta +\beta _{A}\partial
_{a}\alpha ^{A}}{\partial _{n}\epsilon }  \label{PF 4-velocity}
\end{equation}%
that prevents the resulting fluid's velocity from being constrained to
specific kinds of motions. (Let us remind that one $\beta \partial
_{a}\alpha $\ term is enough to ensure the most general velocity field, but
with less than three terms, the Lagrangian coordinates interpretation of the 
$\alpha ^{A}$'s is lost [10].) To be complete, let us also remind that one
usually defines the specific energy $\Pi $ by%
\begin{equation}
\epsilon =n\left( 1+\Pi \right) \text{ \ \ ---\TEXTsymbol{>} \ \ }%
P=n^{2}\partial _{n}\Pi  \label{PF specific energy}
\end{equation}%
and that the PF is endowed with a barotropic equation of state $P\left(
\epsilon \right) $ if and only if $\partial _{s}\epsilon =0$.

\qquad

\noindent \noindent \textbf{III.4.b -- A family of Brown like Lagrangians}

\noindent \noindent Since two matter Lagrangians differing by a divergence
term return the same stress tensor, one can contemplate, whatever the
constant $h$ 
\begin{eqnarray}
L_{PF}^{\prime } &=&L_{PF}-h\nabla _{a}\left[ \left( \phi +s\theta \right)
j^{a}\right]  \label{Brown PF Lagr alt} \\
&=&-\epsilon +j^{a}\left[ \left( 1-h\right) \left( \partial _{a}\phi
+s\partial _{a}\theta \right) -h\theta \partial _{a}s+\beta _{A}\partial
_{a}\alpha ^{A}\right] -h\left( \phi +s\theta \right) \nabla _{a}j^{a} 
\notag
\end{eqnarray}%
as a worth alternative to (\ref{Brown PF Lagr}), as long as GR gravity is
concerned. On the other hand, the matter Lagrangian entering itself the $q$%
-ER equations, the $L_{PF}^{\prime }$ filled $q$-ER equations will be
explicitely $h$ dependent.

It is worth to wonder if some specific choice of $h$ makes (\ref{Brown PF
Lagr alt}) a relevant PF Lagrangian in the $q$-ER gravity framework.

\qquad

\noindent \noindent \textbf{III.4.c -- }$q$\textbf{-ER Matter equations from
the Brown like Lagrangian}

\noindent \noindent Using (\ref{Brown PF Lagr alt}), let us write the $q$-ER
equations. Using (\ref{Brown PF Lagr alt}) for $L_{m}$ in (\ref{q-ER action}%
) and varying w.r.t. the dynamical fields, one finds that the (\ref{1})-(\ref%
{7}) system is replaced by 
\begin{subequations}
\begin{gather}
E_{ab}=\frac{sq}{2\left( 1-q\right) }H^{-\frac{1}{q}}\left[ \left( \epsilon
+n^{2}\partial _{n}\left( \frac{\epsilon }{n}\right) \right)
u_{a}u_{b}+n^{2}\partial _{n}\left( \frac{\epsilon }{n}\right) g_{ab}\right]
+\frac{1}{H}\left( \nabla _{a}\partial _{b}H-g_{ab}\square H\right)
\label{1'} \\
\left( \partial _{n}\epsilon \right) u_{a}=-\partial _{a}\phi -s\partial
_{a}\theta -\beta _{A}\partial _{a}\alpha ^{A}-h\frac{q-1}{q}\left( \phi
+s\theta \right) \partial _{a}\ln H  \label{2'} \\
H^{h\frac{q-1}{q}}\partial _{s}\epsilon =nu^{a}\partial _{a}\left( H^{h\frac{%
q-1}{q}}\theta \right)  \label{3'} \\
\nabla _{a}\left( H^{\left( 1-h\right) \frac{q-1}{q}}nu^{a}\right) =0
\label{4'} \\
u^{a}\partial _{a}s=0  \label{5'} \\
u^{a}\partial _{a}\alpha ^{A}=0  \label{6'} \\
u^{a}\partial _{a}\left( H^{h\frac{q-1}{q}}\beta _{A}\right) =0  \label{7'}
\end{gather}%
Contracting (\ref{2'}) with $u^{a}$\ yields, using (\ref{6'}) and (\ref{5'}) 
\end{subequations}
\begin{equation}
H^{h\frac{q-1}{q}}\partial _{n}\epsilon =u^{a}\partial _{a}\left[ H^{h\frac{%
q-1}{q}}\left( \phi +s\theta \right) \right] .  \label{(2') contracted}
\end{equation}%
As expected, (\ref{q-ER field eq}) is recovered in the (\ref{1'}) form. It
returns the expressions given in (\ref{GR Einstein eq}) for the stress
tensor and the pressure, in terms of $\epsilon $ and $n$. In any case, the
GR equations (\ref{5}) and (\ref{6}) are recovered. The (\ref{2}), (\ref{3}%
), and (\ref{7}) (and (\ref{(2) contracted})) are recovered if one chooses $%
h=0$ (ie, the original Brown Lagrangian). However, (\ref{4}) is not. On the
other hand, recovering (\ref{4}) is achieved choosing $h=1$, but no longer
are (\ref{2}), (\ref{3}) and (\ref{7}) (and (\ref{(2) contracted})). The non
recovering of the whole (\ref{2})-(\ref{7}) system (for non constant $H$'s)
results from the non conservation of the stress tensor.

Let us define 
\begin{equation}
V=L_{PF}^{\prime }-\frac{q}{2\left( q-1\right) }T  \label{non ID indic}
\end{equation}%
the vanishing of which meaning ID. If not vanishing, (\ref{non ID indic})
characterizes how much the PF Lagrangian (\ref{Brown PF Lagr alt}) departs
from the ID condition. Accordingly, let us name it the "non ID quantizer".
Inserting (\ref{2'})-(\ref{7'}) in (\ref{Brown PF Lagr alt}) yields the
on-shell expression%
\begin{equation}
V=\left( \frac{q}{2\left( q-1\right) }-h\right) \epsilon -\left( \frac{q+2}{%
2\left( q-1\right) }+h\right) P.  \label{V on shell}
\end{equation}%
Discarding the case of a very specific linear barotropic PF\footnote{%
Let us also remind that the pressure inside a linear barotropic PF never
vanishes, a property that prevents the existence of stars.}, ID requires%
\begin{equation}
q=-1\text{ \ \ and \ \ }h=\frac{1}{4}  \label{exact ID for PF}
\end{equation}%
that fixes both the theory ($q=-1$) and the PF Lagrangian. Thence, for any ($%
q\neq -1$)-ER, that includes the original ER, the ID condition \textit{cannot%
} be \textit{exactly} satisfied.

\qquad

\noindent \noindent \textbf{III.4.d -- A 1PN Brown like Lagrangian for }$q$%
\textbf{-ER}

\noindent \noindent Having weak field astronomical applications in mind, and
since $P$ is a $1/c^{2}$\ term in the PN context [15], one will say that ID
is complied "up to 1PN order" if $V$ is limited to its pressure term, ie if
the $\epsilon $'s coefficient in (\ref{V on shell}) vanishes. This yields%
\begin{equation}
h=\frac{q}{2\left( q-1\right) }\text{ \ \ ---\TEXTsymbol{>} \ \ }V=-\frac{q+1%
}{q-1}P.  \label{h for 1PN ID}
\end{equation}%
Assuming this choice, the scalar field is only 1PN sourced, in such a way
that solutions having $\partial H=O\left( P\right) $ can be contemplated.\
The PF Lagrangian that allows 1PN ID fulfillment for $q$-ER gravity is then%
\begin{eqnarray}
L_{PF}^{\prime } &=&-\epsilon +j^{a}\left( \partial _{a}\phi +s\partial
_{a}\theta +\beta _{A}\partial _{a}\alpha ^{A}\right) -\frac{q}{2\left(
q-1\right) }\nabla _{a}\left[ \left( \phi +s\theta \right) j^{a}\right]
\label{q-ER PF Lagr ID 2} \\
&=&-\epsilon +j^{a}\left[ \frac{q-2}{2\left( q-1\right) }\left( \partial
_{a}\phi +s\partial _{a}\theta \right) -\frac{q}{2\left( q-1\right) }\theta
\partial _{a}s+\beta _{A}\partial _{a}\alpha ^{A}\right] -\frac{q}{2\left(
q-1\right) }\left( \phi +s\theta \right) \nabla _{a}j^{a}  \notag
\end{eqnarray}%
Let us remark that the baryonic number conservation equation (\ref{4'})
reads, with (\ref{h for 1PN ID}) inserted%
\begin{equation}
\nabla _{a}\left( H^{\frac{q-2}{2q}}nu^{a}\right) =0
\label{(4') with 1PN ID}
\end{equation}%
The baryonic mass is then exactly conserved for $q=2$ only. It is no longer
if $q\neq 2$, but there is a "$q$-modified baryonic number", defined by%
\begin{equation}
N_{q}=H^{\frac{q-2}{2q}}n  \label{q-mod bar number}
\end{equation}%
which is exactly conserved.

Let us note that in the $q=2$ case, (\ref{q-ER PF Lagr ID 2}) simplifies into%
\begin{equation}
L_{PF}^{\prime }=-\epsilon +j^{a}\left( -\theta \partial _{a}s+\beta
_{A}\partial _{a}\alpha ^{A}\right) -\left( \phi +s\theta \right) \nabla
_{a}j^{a}  \label{2-ER PF Lagr ID}
\end{equation}%
while the $2$-modified baryonic number identifies with the usual baryonic
number. This last points fits well with the fact that the (\ref{free part
Lagr}) Lagrangian fills the ID requirement for $q=2$, and only in this case.

\qquad

\noindent \noindent \textbf{IV -- The }$q$\textbf{-ER's Poisson equation}

\noindent \noindent With the previous proposal for the PF description, one
can consider the Poisson equation that emerges in the $q$-ER framework. Let
us consider the weak field case. Using quasi cartesian coordinates, one has%
\begin{equation}
\left\{ 
\begin{array}{l}
g_{ab}=m_{ab}+h_{ab} \\ 
\Phi =1+p%
\end{array}%
\right. \text{ \ \ with \ \ }\left\{ 
\begin{array}{l}
\left( m_{ab}\right) =diag\left( -1,+1,+1,+1\right) \\ 
\left\vert h_{ab}\right\vert \text{ \ \ and \ }\left\vert p\right\vert <<1%
\end{array}%
\right. .  \label{weak field cond}
\end{equation}%
Let us besides require "Newtonian conditions", which means vanishingly small
(1) time variations of the fields ($\partial _{0}\simeq 0$), (2) motions ($%
\left( u^{a}\right) \simeq \left( 1,0,0,0\right) $), and (3) $P/\epsilon $\
ratio. Using (\ref{4'}), (\ref{5'}), (\ref{6'}), (\ref{(2') contracted}), (%
\ref{h for 1PN ID}) and the pressure expression in (\ref{GR Einstein eq}),
the (\ref{q-ER PF Lagr ID 2}) on-shell Lagrangian yields%
\begin{equation}
L_{PF}^{\prime }=-\frac{q}{2\left( q-1\right) }\epsilon +O\left( P\right) .
\label{on shell PF Lagr ID Newt}
\end{equation}%
The $\left( 00\right) $ component of (\ref{1'}) reads, at the same order%
\begin{equation}
2E_{00}=\frac{sq}{1-q}H^{-\frac{1}{q}}\epsilon  \label{Newtonian (00) eq}
\end{equation}%
with%
\begin{equation}
2E_{00}=\partial _{i}\partial _{k}h_{ik}-\Delta h_{kk}.  \label{Newt E(00)}
\end{equation}%
where $\Delta $\ is the Euclidean Laplacian operator. Now, the $\left(
ij\right) $ component of (\ref{1'}) reads, considering solutions such that $%
\partial H\sim P$%
\begin{equation}
E_{ij}=O\left( P\right)  \label{Newtonian (ij) eq}
\end{equation}%
with%
\begin{equation}
2E_{ij}=\partial _{i}\partial _{k}h_{jk}+\partial _{j}\partial
_{k}h_{ik}-\delta _{ij}\partial _{k}\partial _{l}h_{kl}-\Delta h_{ij}+\left(
\partial _{i}\partial _{j}-\delta _{ij}\Delta \right) h_{00}+\left( \delta
_{ij}\Delta -\partial _{i}\partial _{j}\right) h_{kk}.  \label{Newt E(ij)}
\end{equation}%
At the required order, (\ref{Newtonian (ij) eq}) admits the $%
h_{ij}=h_{00}\delta _{ij}$\ solution, which yields, inserting in (\ref%
{Newtonian (00) eq})%
\begin{equation}
\Delta h_{00}=-\frac{qs}{2\left( 1-q\right) }H^{-\frac{1}{q}}\epsilon .
\label{Poisson like eq}
\end{equation}%
Identifying $h_{00}/2$ to the Newtonian potential $U$, this returns the
Poisson equation, with the expression%
\begin{equation}
16\pi G_{eff}=\frac{qs}{1-q}H^{-\frac{1}{q}}  \label{effective G}
\end{equation}%
for the effective gravitational constant $G_{eff}$. This justifies the (\ref%
{positivity cond}) restriction on the sign.

\qquad

\noindent \noindent \textbf{V -- Spherisymmetry and weak field spherical
stars}

\noindent \noindent Let us specify to spherical spacetimes, with matter
confined in the $r\leq R$ region, which defines a "star" as the source of
the field.

For $r>R$, ie outside the star, one knows that $q$-EST identifies with $0$%
-BD. The general spherisymmetric solution is then the BD Class I solution
(the BD Class II, III and IV solutions being only defined for $\omega <-3/2$%
). For $\omega =0$, it reads, using isotropic coords [16] 
\begin{equation}
ds^{2}=-\left( \frac{r-k}{r+k}\right) ^{\frac{2}{\lambda }}dt^{2}+\left( 1+%
\frac{k}{r}\right) ^{4}\left( \frac{r-k}{r+k}\right) ^{2-2\frac{\Lambda +1}{%
\lambda }}\left( dr^{2}+r^{2}d\Omega ^{2}\right) \text{ \ \ with \ \ }%
\lambda ^{2}=\Lambda ^{2}+\Lambda +1.  \label{(w=0) Brans sol 1}
\end{equation}%
Since changing $\lambda $ for $-\lambda $ is equivalent to change $k$ for $%
-k $, one can choose $\lambda \geq 0$\ with no generality loss. Therefore,
the general case can be written in the two parameters form%
\begin{equation}
ds^{2}=-\left( \frac{r-k}{r+k}\right) ^{\frac{2}{\sqrt{\Lambda ^{2}+\Lambda
+1}}}dt^{2}+X^{2}\left( dr^{2}+r^{2}d\Omega ^{2}\right) \text{ \ \ with \ \ }%
X\left( r\right) =\left( 1+\frac{k}{r}\right) ^{2}\left( \frac{r-k}{r+k}%
\right) ^{1-\frac{\Lambda +1}{\sqrt{\Lambda ^{2}+\Lambda +1}}}
\label{(w=0) Brans sol 2}
\end{equation}%
where $k$ can a priori have both signs. In the weak field region, (\ref%
{(w=0) Brans sol 2}) Taylor expands into%
\begin{equation}
ds^{2}=-\left[ 1-2\frac{m}{r}+2\frac{m^{2}}{r^{2}}+O\left( \frac{1}{r^{3}}%
\right) \right] dt^{2}+\left[ \allowbreak 1+2\gamma _{Edd}\frac{m}{r}+\frac{3%
}{2}\delta _{Edd}\frac{m^{2}}{r^{2}}+O\left( \frac{1}{r^{3}}\right) \right]
\left( dr^{2}+r^{2}d\Omega ^{2}\right)  \label{(w=0) Brans sol exp}
\end{equation}%
with%
\begin{equation}
\left\{ 
\begin{array}{l}
m=\frac{2k}{\sqrt{\Lambda ^{2}+\Lambda +1}} \\ 
\gamma _{Edd}=\Lambda +1 \\ 
\delta _{Edd}=\allowbreak \Lambda ^{2}+\frac{7}{3}\Lambda +1%
\end{array}%
\right. .  \label{(w=0) Brans sol PN param}
\end{equation}%
Thence, contemplating only the possibility of positive gravitational mass
stars, $k$ has now to be positive\footnote{%
Let us remind that for $k>0$, (\ref{(w=0) Brans sol 2}) describes a (vacuum)
wormhole like or a naked singularity sourced spacetime, depending on the $%
\Lambda $\ sign, as it can easily be checked from the areal radius $X\left(
r\right) $ and the sign of $\left( 1-\frac{\Lambda +1}{\sqrt{\Lambda
^{2}+\Lambda +1}}\right) $.}.

\qquad

Let us remind that in $0$-BD, and for a weak field Newtonian star (that
requires $P<<\epsilon $ inside the star), one finds [8]%
\begin{equation}
\Lambda _{BD}=-\frac{1}{2}\text{ \ \ ---\TEXTsymbol{>} \ \ }\gamma _{Edd,BD}=%
\frac{1}{2}\text{ \ \ and \ \ }\delta _{Edd}=\allowbreak \frac{1}{12}
\label{(w=0) BD PN param}
\end{equation}%
which makes the theory unviable w.r.t. experiments. However, this $\Lambda
_{BD}$ value results from the integration of the BD equations inside the
star. Since $q$-ER differs from BD in matter, one expects $\Lambda $ to
differ from $\Lambda _{BD}$, making a priori possible the theory to comply
with the experimental tests to date.

\qquad

For $q$-EST to pass the experimental tests, one needs $\gamma _{Edd}$ to be
close enough to $1$, ie $\Lambda $ to be close enough to $0$ (that
incidentally would imply $\delta _{Edd}$ to be close to $1$ too).

\qquad

\noindent \noindent \textbf{V.1 -- Weak field spherical stars in }$q$\textbf{%
-EST}

\noindent \noindent Let us now specify to the weak field case (\ref{weak
field cond}), but without specifying to PF stars for the moment. The (\ref%
{q-ER renorm field eq})-(\ref{q-ER renorm scalar eq}) equations yield, in
the linear approximation 
\begin{subequations}
\begin{gather}
m^{ce}\left( \partial _{a}\partial _{c}h_{be}+\partial _{b}\partial
_{c}h_{ae}-\partial _{c}\partial _{e}h_{ab}-\partial _{a}\partial
_{b}h_{ce}\right) =16\pi \left( T_{ab}-\frac{1}{2}Tm_{ab}-\frac{q-1}{3q}%
Vm_{ab}\right) +2\partial _{a}\partial _{b}p  \label{lin q-ER Einstein eq} \\
m^{ce}\partial _{c}\partial _{e}p=-\frac{16\pi }{3}\frac{q-1}{q}V
\label{lin q-ER scalar eq}
\end{gather}%
where the $h_{\ast \ast }$'s and $p$ are defined by (\ref{weak field cond}),
and $V$ is the non ID quantifier (\ref{non ID indic}). Outside the star, one
has, from (\ref{(w=0) Brans sol exp}) 
\end{subequations}
\begin{equation}
\left( h_{ab}\right) =2\frac{m}{r}\times diag\left( 1,\gamma _{Edd},\gamma
_{Edd},\gamma _{Edd}\right) \text{ \ \ ---\TEXTsymbol{>} \ \ }h_{ij}=\gamma
_{Edd}h_{00}\delta _{ij}.  \label{weak field outside the star}
\end{equation}%
Switching to spherical coordinates for rewriting the $m^{ce}\partial
_{c}\partial _{e}=m^{ij}\partial _{i}\partial _{j}$ Laplacian operator, and
integrating (\ref{lin q-ER scalar eq}) from the star's center up to an
external point yields 
\begin{equation}
\frac{dp}{dr}\left( r>R\right) =-\frac{16\pi }{3}\frac{q-1}{q}\frac{1}{r^{2}}%
\int_{0}^{R}Vr^{2}dr  \label{lin q-ER scalar eq integr}
\end{equation}%
where the regularity condition at the star's center has been used. This
equation explicits how the GR departure (non constant scalar) is triggered
by the ID violation inside the star. Similarily, the $\left( 00\right) $
component of (\ref{lin q-ER Einstein eq}) integrates into%
\begin{equation}
\frac{dh_{00}}{dr}\left( r>R\right) =-\frac{16\pi }{r^{2}}\int_{0}^{R}\left(
T_{00}+\frac{1}{2}T+\frac{q-1}{3q}V\right) r^{2}dr.
\label{lin q-ER (00) Einstein eq integr}
\end{equation}%
Finally, writing the $i\neq j$ components of (\ref{lin q-ER Einstein eq})
outside the star yields%
\begin{equation}
\left( 1-\gamma _{Edd}\right) \frac{d}{dr}\left( \frac{1}{r}\frac{dh_{00}}{dr%
}\right) =2\frac{d}{dr}\left( \frac{1}{r}\frac{dp}{dr}\right) .
\label{lin q-ER non diag (ij) Einstein eq}
\end{equation}%
(The diagonal spatial components of (\ref{lin q-ER Einstein eq}) carry
nothing new.) Now, inserting (\ref{lin q-ER scalar eq integr}) and (\ref{lin
q-ER (00) Einstein eq integr}) in (\ref{lin q-ER non diag (ij) Einstein eq})
returns%
\begin{equation}
1-\gamma _{Edd}=\frac{2}{3}\left( 1-\frac{1}{q}\right) \frac{%
\int_{0}^{R}Vr^{2}dr}{\int_{0}^{R}\left( T_{00}+\frac{1}{2}T+\frac{q-1}{3q}%
V\right) r^{2}dr}.  \label{q-ER gamma_Edd 1}
\end{equation}%
It gives the $q$-EST's 1PN departure from GR (reminding that, from (\ref%
{(w=0) Brans sol exp}), $\beta _{Edd}=1$). As expected, $\gamma _{Edd}=1$ if 
$V=0$, ie if the star's matter \textit{exactly} complies with the ID
requirement.

\qquad

\noindent \noindent \textbf{V.2 -- PF weak field spherical stars}

\noindent \noindent Specifying to PF stars (meaning reminded just after (\ref%
{weak field cond})), (\ref{q-ER gamma_Edd 1}) rewrites, inserting the (\ref%
{h for 1PN ID}) 1PN requirement and keeping the dominant term only%
\begin{equation}
1-\gamma _{Edd}=-\frac{4}{9}\frac{q+1}{q}\Theta \text{ \ \ where \ \ }%
\left\{ 
\begin{array}{l}
\Theta =\frac{3}{m}\int_{0}^{R}P\times 4\pi r^{2}dr \\ 
m=\int_{0}^{R}\epsilon \times 4\pi r^{2}dr%
\end{array}%
\right. .  \label{q-ER gamma_Edd  if ID}
\end{equation}%
Let us remark that one has exactly $\gamma _{Edd}=1$ for the $q=-1$ theory,
in accordance with the fact that this theory exactly complies with ID (see (%
\ref{exact ID for PF})).

Inserting in (\ref{q-ER gamma_Edd if ID}) the Sun's value of $\Theta $ that
was used in [17], refering to [18], yields%
\begin{equation}
\Theta \simeq 3.43\times 10^{-6}\text{ \ \ ---\TEXTsymbol{>} \ \ }1-\gamma
_{Edd}=-\frac{q+1}{q}\times 1.52\times 10^{-6}.  \label{ER gamma_Edd 3}
\end{equation}%
In the $q=2$ case (original ER theory), (\ref{ER gamma_Edd 3}) gives $\gamma
_{Edd}-1=2.28\times 10^{-6}$, a value that is about one third the value
obtained in [19] (see their eq. (82a))\footnote{%
This factor originates in a confusion between the total pressure, as defined
in [17], and the PF's pressure.}.

\qquad

\noindent \noindent \textbf{V.3 -- On the non conservation of the stress
tensor and its consequences}

\noindent \noindent Since the Lagrangian (\ref{q-ER PF Lagr ID 2}) doesn't
exactly fill the ID condition, it is worth asking how much the stress tensor
conservation is violated inside the weak field PF star previously
considered. It is also worth to ask the way test matter behaves outside the
star in the $q$-EST perspective, ie to discuss the consequences of the test
matter stress tensor non conservation. We examine these points here, and the
way the known contraints on PN parameters justify $q$-ER as an
experimentally viable gravity theory.

The (\ref{q-ER stress tensor non conserv}) equation rewrites, using (\ref%
{weak field cond})%
\begin{equation}
\nabla _{b}T_{a}^{b}=\frac{q-1}{q}\left( L_{PF}^{\prime }\delta
_{a}^{b}-T_{a}^{b}\right) \partial _{b}p.  \label{1PN stress conserv}
\end{equation}

\qquad

\noindent \noindent \textbf{V.3.a -- Inside the star}

\noindent \noindent Integrating (\ref{lin q-ER scalar eq}) inside the star
with (\ref{h for 1PN ID}) and inserting in (\ref{1PN stress conserv}) yields%
\begin{equation}
\frac{dp}{dr}\left( r<R\right) =\frac{4}{3}\frac{q+1}{q}\frac{1}{r^{2}}%
\int_{0}^{r}P\times 4\pi r^{2}dr.  \label{pressure inside}
\end{equation}%
Inserting (\ref{pressure inside}) in (\ref{1PN stress conserv}) yields the
dominant contribution, using (\ref{on shell PF Lagr ID Newt}) and $u^{r}=0$
(static star)%
\begin{equation}
\nabla _{b}T_{a}^{b}=-\frac{2}{3}\frac{q+1}{q}\delta _{a}^{r}\frac{\epsilon 
}{r^{2}}\int_{0}^{r}P\times 4\pi r^{2}dr.  \label{1PN stress conserv 2}
\end{equation}%
As expected, the PF stress tensor is not conserved, since the radial
component of (\ref{1PN stress conserv 2}) doesn't vanish. For an order of
magnitude, using the $\Theta $ quantity defined by (\ref{q-ER gamma_Edd if
ID}), one sees that, neglecting dimensionless numerical factors, this radial
component is of order $\Theta \epsilon m/R^{2}$, since $r$ and $R$ are
comparable inside the star. Since the dominant term of the stress tensor is
the energy density, the typical order of magnitude of terms entering the
stress divergence calculation is expected to be of $\epsilon /R$ order. The
ratio of these two quantities is of order $\Theta m/R$, ie, reintroducing
standard units%
\begin{equation}
\frac{\left( \nabla _{b}T_{a}^{b}\right) _{r}}{\epsilon /R}\sim \Theta \frac{%
Gm}{Rc^{2}}.  \label{1PN stress conserv num}
\end{equation}%
For the Sun, using the (\ref{ER gamma_Edd 3}) $\Theta $ and the
Schwarzschild radius values, this ratio is of order $10^{-11}$. The stress
tensor conservation is then very close to be conserved inside a Sun like
star in the $q$-ER framework, once (\ref{q-ER PF Lagr ID 2}) is chosen as
the relevant PF Lagrangian.

Let us remark that $\nabla _{b}T_{a}^{b}$ exactly vanishes for the $q=-1$
theory, in accordance with the fact that $\left( -1\right) $-ER exactly
complies with ID (see (\ref{exact ID for PF})).

\qquad

\noindent \noindent \textbf{V.3.b -- Outside the star}

\noindent \noindent Inserting (\ref{lin q-ER scalar eq integr}) and (\ref{h
for 1PN ID}) in (\ref{1PN stress conserv}) yields, using the $\Theta $\
quantity (\ref{q-ER gamma_Edd if ID})%
\begin{equation}
\nabla _{b}T_{a}^{b}=\frac{4}{9}\frac{q^{2}-1}{q^{2}}\left( L_{m}\delta
_{a}^{r}-T_{a}^{r}\right) \frac{m\Theta }{r^{2}}.  \label{out 1}
\end{equation}%
Considering PF balls, the PF Lagrangian (\ref{on shell PF Lagr ID Newt}) and
stress tensor in (\ref{GR Einstein eq}) yield, since $P<<\epsilon $ and $%
u^{r}<<1$ for planetary like motions 
\begin{equation}
\frac{\nabla _{b}T_{a}^{b}}{\epsilon /r}=-\frac{2}{9}\frac{q+1}{q}\delta
_{a}^{r}\Theta \frac{Gm}{Rc^{2}}\times \frac{R}{r}  \label{out 3}
\end{equation}%
where standard units have been reintroduced. This results in orders of
magnitude damped down by the $R/r$ factor w.r.t. (\ref{1PN stress conserv
num}). Applied to our Solar System, with $R/r\sim 10^{-2}$ for inner
planets, this yields $\sim 10^{-13}$ relative deviations from the conserved
case.

Considering propagation of electromagnetic waves, (\ref{out 1}) is still
relevant, but with the electromagnetic Lagrangian (\ref{em ID test}) and
stress tensor (\ref{em stress tensor}) expressions. This yields, for its
dominant term, in the radiative case%
\begin{equation}
\nabla _{b}\left( 4F_{ac}F^{bc}-\delta
_{a}^{b}g^{pq}g^{ce}F_{pc}F_{qe}\right) =-\frac{16}{9}\frac{q^{2}-1}{q^{2}}%
\frac{m\Theta }{r^{2}}m^{ce}F_{ac}F_{re}  \label{out 4}
\end{equation}%
that provides how much electromagnetic waves' behaviour departs from the
conservative case.

\qquad

\noindent \noindent \textbf{V.3.c -- Testing }$q$\textbf{-ER in the Solar
System}

\noindent \noindent The $\left( 1-\gamma _{Edd}\right) $ value obtained in (%
\ref{ER gamma_Edd 3}) is about one order of magnitude below currently known
observational constraints [9]. This suggests that $q$-ER gravity is an
experimentally viable alternative to GR, at least w.r.t. Solar System like
experiments. However, one could argue that a complete and coherent testing
of $q$-ER involving Solar System observational data would first require
reconsidering the description of planetary motions and propagation of light
in this framework. Indeed, the stress tensor non conservation conflicts the
usual PPN treatment, that is grounded on the stress tensor conservation
assumption, that notably implies the null geodesic moving of light [15].
Concerning the propagation of light, (but only its propagation), there is no
specific non conservation effects to worry about, since light still moves
along null geodesics in $q$-ER, as it is shown in the Appendix (which
generalizes the result obtained in the $q=2$ case, see [7]). The case is not
so simple for planetary motions, that would require developping (\ref{out 1}%
) in the case of PF balls, since massive bodies of the Solar System are
modelized this way in the PPN approach [15]. This is not the purpose of the
current paper to develop this side of the work, but an order of magnitude of
the effect can nevertheless be attempted. Indeed, the spatial distance
spanned over a complete revolution by a planet is $\sim 2\pi r$, where $r$
is the typical size of its orbit. From (\ref{out 3}), the expected position
shift induced by the non conservation term should then be of order, using
the Solar $\Theta $\ value of (\ref{ER gamma_Edd 3})%
\begin{equation}
2\pi r\times \Theta \frac{Gm}{Rc^{2}}\times \frac{R}{r}\sim 10^{-5}\times 
\frac{2Gm}{c^{2}}.  \label{q-ER departure from geod}
\end{equation}%
In the Solar System, (\ref{q-ER departure from geod}) is a centrimetric
quantity since $\frac{2Gm}{c^{2}}\sim 3\ km$ for the Sun, and just reaches
the meter level after some dozen revolutions. This suggests that the PPN
constraint on $\gamma _{Edd}$ should be applicable even in this $q$-ER
framework. Nonetheless, it is clear that a precise study specifically
adressing this issue would be welcome.

\qquad

\noindent \noindent \textbf{VI -- Discussion and conclusion}

\noindent \noindent In this paper, one reconsidered the ER (purely tensor)
gravity framework, but encompassing it in a one parameter ($q$) family of
theories. All these $q$-ER theories share the very idea of original ER (that
corresponds to the $q=2$ case): for making sense, the underlying Lagrangian
deeply needs the contribution of a non vanishing matter Lagrangian.
Nonetheless, these theories admit an ST reformulation, in which considering
vacuum spacetime regions, and even fully vacuum solutions, makes sense. A
specific feature of these theories is that the scalar of the ST formulation
is sourced not only by the (trace of the) stress tensor, but also by the
matter Lagrangian itself. This makes the theory non invariant by the
addition of a divergence term to the matter Lagrangian, a deep difference
w.r.t. usual gravity theories.

One has shown that the ST version of all these theories identify with a
single theory in the vacuum regions of spacetime, that is nothing but the BD
theory with $\omega =0$. One the other hand, all these theories differ each
other (and differ from any BD theory) inside matter. It turns out that,
whatever $q$, these theories admit all the GR solutions if the matter
Lagrangian satisfies a specific condition (ID). One has shown that $\left( 
\frac{1}{q}+1\right) $-$g^{\ast \ast }$-homogeneous Lagrangians
systematically yield ID, and that for such Lagrangians, non GR solutions are
conformally associated by pairs.

Having astronomical applications in mind, one has focused on PF matter
contents. A shortcoming of PFs is that, while the stress tensor is
well-known, the structure of the Lagrangian from which it proceeds lacks a
clear physically grounded justification. A Lagrangian is neverthelees known
from long that does the job w.r.t. the stress tensor. One has shown that an
appropriate modification of this Lagrangian, that doesn't modify the stress
tensor, can be fixed in a way that allows weak field stars to departs from
ID by 1PN terms only. In the case of the Sun, the $\left( 1-\gamma
_{Edd}\right) $ numerical value is one order of magnitude below the
departure allowed by Solar System observations. At first sight, this
suggests that $q$-ER could be worth alternatives to GR. However, a firm
claim would first require a close examination on the way the $\gamma _{Edd}$
PN parameter measurement is related to PF balls motion in this framework
(planets being modelized as PF balls in PPN gravity). Indeed, the PF Euler
equations are no longer exactly recovered in $q$-ER gravity. A study
specifying the way the $q$-ER stress tensor non conservation impacts the
link between the Eddington parameter values and their measurements would be
welcome.

It has been marginally remarked (see (\ref{exact ID for PF})) that the $q=-1$
theory allow exact ID for PFs. The point could seem advantageous w.r.t. the $%
q=2$ original ER, that only allows 1PN ID for PFs. On the other hand, the
free particle Lagrangian doesn't comply with ID at all, even in its on-shell
version, as shown by (\ref{free part ID test}), while it does for $q=2$. Let
us also remark that in the pressureless case, ie for dust which is nothing
but a collection of free particles, the on-shell PF Lagrangian (\ref{h for
1PN ID}) exactly vanishes. These two descriptions of a gas of free particles
are then fully coherent in the original ER framework.

\qquad

\noindent \noindent \textbf{Acknowledgements}

\noindent \noindent The author thanks Olivier Minazzoli (Bureau des Affaires
Spatiales, 2 rue du Gabian, Monaco 98000, Monaco), and Aur\'{e}lien Hess
(LTE Observatoire de Paris, Universit\'{e} PSL, Sorbonne Universit\'{e},
Universit\'{e} de Lille, LNE, CNRS, 61 Avenue de l'Observatoire, 75014
Paris, France) for fruitful comments and discussions.

\qquad

\noindent \noindent \textbf{Appendix: on light paths in }$q$\textbf{-ER}

\noindent \noindent The (\ref{q-ER renorm stress tensor non conserv}) non
conservation law yields, with (\ref{em Lagr})\ and (\ref{em stress tensor}) 
\begin{equation}
\partial _{b}\left[ \sqrt{-g}\left( g_{ce}F^{ac}F^{be}-\frac{1}{4}%
g^{ab}F_{ce}F^{ce}\right) \right] =-\frac{q-1}{q}\sqrt{-g}%
g_{be}F^{ab}F^{ce}\partial _{c}\ln H.  \label{App2 non conserv em}
\end{equation}%
The GOA consists in looking for solutions of the form [20]%
\begin{equation}
A^{c}=\func{Re}\left\{ \left[ a^{c}+O\left( \epsilon \right) \right] \exp
\left( i\frac{\theta }{\varepsilon }\right) \right\} \text{ \ \ ---%
\TEXTsymbol{>} \ \ }\partial ^{a}A^{c}=-\frac{1}{\varepsilon }k^{a}\left( 
\widehat{a}^{c}\cos \chi +\widetilde{a}^{c}\sin \chi \right) +O\left(
\varepsilon ^{0}\right)  \label{App2 for GO}
\end{equation}%
where $\theta $ is a scalar field, $\varepsilon $ a vanishingly small
parameter in such a way that the phase $\chi =\frac{\theta }{\varepsilon }$
is rapidly varying w.r.t. any other involved scale entering the problem, $%
a^{\ast }=\widetilde{a}^{\ast }+i\widehat{a}^{\ast }$\ is the complex
amplitude vector (that provides the potential vector polarization besides
the electromagnetic wave amplitude), and%
\begin{equation}
k=\partial \theta  \label{App2 k def}
\end{equation}%
a (real) vector, named wave vector. Both $\theta $ and $a^{\ast }$ are $%
\varepsilon $ independent quantities. The (\ref{App2 non conserv em}) rhs is
an $O\left( \varepsilon ^{0}\right) $\ quantity. From (\ref{App2 for GO}), $%
F^{\ast \ast }$ is an $\varepsilon ^{-1}$ quantity. Deriving $F^{\ast \ast }$
introduces terms involving a new $\varepsilon ^{-1}$\ factor, while deriving
the metric doesn't. Thence, (\ref{App2 non conserv em}) yields%
\begin{equation}
g_{ce}\partial _{b}\left( F^{ac}F^{be}\right) -\frac{1}{2}%
g^{ab}F_{ce}\partial _{b}F^{ce}=O\left( \varepsilon ^{-2}\right)
\label{App2 a}
\end{equation}%
with, from (\ref{App2 for GO})%
\begin{equation}
F^{ac}=\frac{1}{\varepsilon }\left[ \left( k^{c}\widehat{a}^{a}-k^{a}%
\widehat{a}^{c}\right) \cos \chi +\left( k^{c}\widetilde{a}^{a}-k^{a}%
\widetilde{a}^{c}\right) \sin \chi \right] +O\left( \varepsilon ^{0}\right) .
\label{App2 b}
\end{equation}%
As usual, let us demand $A^{\ast }$ to satisfy the usual divergenceless
gauge condition%
\begin{equation}
\nabla _{c}A^{c}=0\text{ \ \ \TEXTsymbol{<}---\TEXTsymbol{>} \ \ }\partial
_{c}\left( \sqrt{-g}A^{c}\right) =0\text{ \ \ ---\TEXTsymbol{>} \ \ }%
\partial _{c}A^{c}=O\left( \varepsilon ^{0}\right)  \label{App2 gauge}
\end{equation}%
that is always possible thanks to the fact that $A^{\ast }$ is defined up to
a gradient from (\ref{em Lagr}). This yields, thanks to (\ref{App2 for GO}),
and involving the same approximations%
\begin{equation}
\frac{1}{\varepsilon }k_{c}\left( \widehat{a}^{c}\cos \chi +\widetilde{a}%
^{c}\sin \chi \right) =O\left( \varepsilon ^{0}\right) \text{ \ \ ---%
\TEXTsymbol{>} \ \ }k_{c}\left( \widehat{a}^{c}\cos \chi +\widetilde{a}%
^{c}\sin \chi \right) =0  \label{App2 c}
\end{equation}%
Les us stress that deriving (\ref{App2 c}) wrt $x^{e}$ yields (still
retaining the dominant term)%
\begin{equation}
k_{c}\left( \widetilde{a}^{c}\cos \chi -\widehat{a}^{c}\sin \chi \right) =0.
\label{App2 d}
\end{equation}%
Inserting (\ref{App2 b}) into (\ref{App2 a}) yields, retaining the dominant (%
$\varepsilon ^{-3}$) term and using (\ref{App2 c}) and (\ref{App2 d})%
\begin{equation}
k_{e}k^{e}\sin \left[ 2\left( \psi +\chi \right) \right] =0\text{ \ \ ---%
\TEXTsymbol{>} \ \ }k_{e}k^{e}=0  \label{App2 null k}
\end{equation}%
where $\psi $\ is the $a^{\ast }$'s argument. Thence, deriving (\ref{App2
null k}) yields, since $\left( \nabla _{a}\partial _{b}-\nabla _{b}\partial
_{a}\right) \theta =0$%
\begin{equation}
k^{e}\nabla _{e}k_{c}=0.  \label{App2 k geod}
\end{equation}%
Thence, (\ref{App2 null k}) and (\ref{App2 k geod}) show that, as in GR,
light rays follow null geodesics in $q$-ER, as they do in the $q=2$ case.
This doesn't come as a surprise, since the scalar contribution in (\ref{App2
non conserv em}), along with the $q$ value, carry negligeable contributions
at the GOA. The departure from the GR case then concerns only the energy and
polarization transfer along the light path (besides post-GOA effects on the
light path itself).

Using (\ref{App2 for GO}) and (\ref{App2 b}), it can be directly checked
that the dominant ($\varepsilon ^{-2}$) term of $g^{ac}g^{be}F_{ab}\partial
_{c}A_{e}=F^{ce}\partial _{c}A_{e}$ vanishes in the GOA, thanks to (\ref%
{App2 c}) and (\ref{App2 null k}).

\qquad

\noindent \noindent \textbf{References}

\noindent \noindent \lbrack 1] H. Ludwig, O. Minazzoli, S. Capozziello,
Phys. Lett. B 751, 576 (2015).

\noindent \noindent \lbrack 2] O. Minazzoli, A. Hees, Phys. Rev. D 94,
064038 (2016).

\noindent \noindent \lbrack 3] O. Minazzoli, Phys. Rev. D 98, 124020 (2018).

\noindent \noindent \lbrack 4] D. Arruga, O. Minazzoli, Eur. Phys. J. C 81,
1027 (2021).

\noindent \noindent \lbrack 5] D. Arruga, O. Rousselle, O. Minazzoli, Phys.
Rev.\ D 103, 024034 (2021).

\noindent \noindent \lbrack 6] O. Minazzoli, E. Santos, Eur. Phys. J. C 81,
640 (2021).

\noindent \noindent \lbrack 7] T. Chehab, O. Minazzoli, A. Hees, Class.
Quantum Grav. 43, 015025 (2026).

\noindent \noindent \lbrack 8] C. Brans, R. H. Dicke, Phys. Rev. 124, 925
(1961).

\noindent \noindent \lbrack 9] C.M. Will, Living Rev. Relat. 17, 4 (2014).

\noindent \noindent \lbrack 10] J.\ D.\ Brown, Class. Quantum Grav. 10, 1579
(1993).

\noindent \noindent \lbrack 11] O. Minazzoli, M. Wavasseur, Eur. Phys. J. C
85, 474 (2025).

\noindent \noindent \lbrack 12] J. B. Griffiths and J. Podolsk\'{y}, Exact
Space-Times in Einstein's General Relativity (Cambridge University Press,
Cambridge, England, 2009).

\noindent \noindent \lbrack 13] A.\ G. Agnese, M. La Camera, Phys. Rev. D
51, 2011 (1995).

\noindent \noindent \lbrack 14] R. M. Wald, General Relativity (Chicago
University Press, Chicago, 1984).

\noindent \noindent \lbrack 15] C. M.Will, Theory and Experiment in
Gravitational Physics (Cambridge University Press, Cambridge, England, 2018).

\noindent \noindent \lbrack 16] C. H. Brans, Phys. Rev. 125, 2194 (1962).

\noindent \noindent \lbrack 17] B. Chauvineau, H. K. Nguyen, Phys. Lett. B
855, 138803 (2024).

\noindent \noindent \lbrack 18] J. Christensen-Dalsgaard et al, Science 272,
1286 (1996).

\noindent \noindent \lbrack 19] T. Chehab, O. Minazzoli, Phys. Rev. D 114,
024012 (2026).

\noindent \noindent \lbrack 20] C.W. Misner, K.S. Thorne, J.A. Wheeler,
Gravitation, Freeman, San Francisco (1973).

\end{document}